\documentclass[nofootinbib,aps,amssymb,floatfix,superscriptaddress,preprintnumbers]{revtex4}
\usepackage{inputenc}
\usepackage{epsfig,float}
\usepackage{amsfonts}
\usepackage{amsmath}
\usepackage{subcaption}
\newcommand{\ket}[1]{\ensuremath{\left|#1\right\rangle}\xspace}

\usepackage[svgnames]{xcolor}
\usepackage{xspace}

\restylefloat{figure}
\usepackage{graphicx}
\usepackage[colorlinks=true, allcolors=blue]{hyperref}
\usepackage[sort&compress]{natbib}
\usepackage{bm}
\usepackage{slashed}

\begin{document}
\title{Backward DVCS in a Sullivan process}

\author{Abigail R. Castro}
\affiliation{Irfu, CEA, Université Paris-Saclay, 91191 Gif-sur-Yvette, France}
\author{C\'edric Mezrag}
\affiliation{Irfu, CEA, Université Paris-Saclay, 91191 Gif-sur-Yvette, France}
\author{Jose M. Morgado Chávez}
\affiliation{Irfu, CEA, Université Paris-Saclay, 91191 Gif-sur-Yvette, France}
\affiliation{Dpto. de Física Teórica and IFIC, Universitat de València and CSIC, 46100 Valencia, Spain}
\author{Bernard~Pire}
\affiliation{CPHT, CNRS, \'{E}cole polytechnique, I.P. Paris,  91128 Palaiseau, France}

\begin{abstract}
Mesons' internal structure and dynamics may be accessed through hard exclusive electroproduction processes such as deeply virtual Compton scattering in both near forward and near backward kinematics. With the help of the Sullivan process which allows us to use a nucleon target as a quasi-real $\pi$ meson emitter, we study backward scattering in the framework of collinear QCD factorization where pion-to-photon transition distribution amplitudes describe the photon content of the $\pi$ meson.
We present a model of these TDAs based on the overlap of lightfront wave functions  primarily developed for generalized parton distributions, using a previously developed pion lightfront wave function and deriving a new model for the lightfront wave functions of the photon.
This leads us to an estimate of the cross-sections for JLab energies. We conclude that deeply virtual $ep\rightarrow e\gamma M n$ processes, in backward kinematics, may be experimentally discovered in the near future.
\end{abstract}

\maketitle


\section{Introduction}

Hard exclusive leptoproduction processes in the so-called Bjorken regime -- where the virtual photon scale $Q^2$ and the process squared energy $s$ are commensurate and much larger than the squared momentum transfer $-t$ of the target -- have been one of the main subjects of interest to hadron physics in the last 30 years. This interest has been caused by the discovery of the factorization of the QCD amplitude in terms of generalized parton distributions (GPDs) \cite{Mueller:1998fv,Ji:1996ek,Ji:1996nm,Radyushkin:1996ru,Radyushkin:1997ki,Collins:1998be} which have proven to be a unique way to perform the quark and gluon tomography of hadrons \cite{Burkardt:2000za,Ralston:2001xs,Diehl:2002he}. For reviews, see \cite{Diehl:2003ny,Belitsky:2005qn}.

The studies of exclusive processes have been mostly focused on leptoproduction on nucleon and nuclei targets, thus giving access to nucleon tomography. Complementary studies in electron-positron facilities \cite{Belle:2015oin, Kumano:2017lhr} allow to probe the quark and gluon content of mesons in reactions such as $e\gamma \rightarrow e M \overline{M}$, thanks to the factorization of their amplitudes through generalized distribution amplitudes \cite{Diehl:1998dk,Polyakov:1998ze,Diehl:2000uv,Pire:2002ut}, which are crossed matrix elements of the same operators as in GPDs. Moreover, recent efforts suggested that (forward) deeply virtual Compton scattering (DVCS) on a virtual pion target might be achievable \cite{Amrath:2008vx,Chavez:2021llq,Chavez:2021koz}, through the so-called Sullivan process \cite{Sullivan:1971kd}. The Sullivan process has been already exploited to extract the pion form factor \cite{NA7:1986vav,JeffersonLab:2008jve}, parton distribution functions (PDFs) \cite{Barry:2018ort,Barry:2021osv} and even transverse momentum dependent PDFs \cite{Barry:2023qqh}. Virtuality effects are usually neglected \cite{Qin:2017lcd}, although they may trigger modifications of the formalism with respect to on-shell targets \cite{Broniowski:2022iip,Shastry:2023fnc}. Note that deep virtual meson production more complicated processes might be also measurable through the Sullivan process \cite{Hatta:2025ryj,Son:2024uxa}. The theoretical counterparts of these experimental and phenomenological activities are the numerous efforts to explore the pion structure, either from lattice QCD \cite{Bali:2017ude,Joo:2019bzr,Alexandrou:2020gxs,Holligan:2023rex,Ding:2024saz} or continuum methods \cite{Mezrag:2014jka,Mezrag:2016hnp,Chouika:2017rzs,Ding:2019lwe,Cui:2021mom,Lu:2023yna,Raya:2024glv,Son:2024uet}.

Backward reactions have recently been the subject of a renewed interest \cite{Gayoso:2021rzj}, in particular in the context of a factorized description of their amplitudes in terms of nucleon to meson or nucleon to photon transition distribution amplitudes (TDAs) \cite{Pire:2004ie, Pire:2021hbl}. First data have been analyzed in this framework \cite{CLAS:2017rgp, JeffersonLabFp:2019gpp, Pire:2022kwu} and are quite encouraging. While these TDAs with non-zero baryonic number are hadronic matrix elements of a three quark operator on the light-cone, mesonic TDAs such as those appearing in a meson to photon~\cite{Pire:2004ie} transition are defined as Fourier transforms of quark-antiquark operators 
very similarly as generalized parton distributions (GPDs). The information contained in these meson-to-photon TDAs pins down the non-perturbative photon content of the mesons, and so provides an interesting way to shed light on how QCD and QED intertwine in the non-perturbative sector. This is complementary to GPDs, encoding the transverse extension of meson~\cite{Burkardt:2002hr, Ralston:2001xs, Diehl:2002he}. Extracting these non-perturbative objects from experimental data is a challenge~\cite{Lansberg:2006fv} that we take up here.

In this paper, we firstly derive a model for the pion to photon TDAs, using lightfront wave functions (LFWF) for the meson and the photon, and calculating their overlap in the so-called Dokshitzer-Gribov-Lipatov-Altarelli-Parisi (DGAP) region. We then use the method developed in \cite{Chouika:2017dhe, Chavez:2021llq, DallOlio:2024vjv} to deduce a model for both the DGLAP and the Efremov-Radyushkin-Brodsky-Lepage (ERBL) regions, this model satisfies all required properties of the TDAs, including polynomiality. We then calculate the amplitude for the electroproduction of a real photon off a $\pi^+$ meson emitted by the nucleon target in the backward kinematic region. This process may occur with various final states. We consider both 
\begin{align}
    \gamma^{\ast}(q)+\pi^{+}(p_\pi)   & \rightarrow \gamma (q') +  \pi^{+}(p'_{\pi}) \,, \\
    \gamma^{\ast}(q)+\pi^{+}(p_{\pi}) & \rightarrow \gamma(q') + \rho^{+}(p'_{\rho}),
\end{align}
where $Q^2 = -q^{2} \gg -u = -(q'-p_\pi)^2$, which probe different TDAs because of the different quantum numbers of the final meson. We then evaluate the cross-sections at JLab 22 kinematics energies and briefly discuss the feasibility of the experiment.

\section{Modelling meson to photon TDAs with lightfront wave functions}\label{sec:TDA}

For starting, let us recall that there are four leading twist $\pi - \gamma$ TDAs: one vector, one axial and two transversity TDAs. In a $\gamma^{\ast} \pi \to \gamma \pi$ process, only the axial quark TDA $A^{\pi\gamma}$ contributes. It is defined\footnote{For simplicity, we omit the Wilson lines required to ensure gauge invariance of the definition.} as~\cite{Pire:2004ie,Lansberg:2006fv}
\begin{equation}\label{eq:AxialTDADef}
\frac{e}{f_{\pi}}\left(\epsilon\cdot\Delta\right)A^{\pi\gamma}(x,\xi,u,\mu)=\frac{1}{2}\int\frac{dz^{-}}{2\pi}e^{ix P^{+}z^{-}}\left.\langle\gamma(q',\epsilon) |\bar{\psi}_{d}(-z/2)\gamma^{+}\gamma_{5}\psi_{u}(z/2)|\pi(p_{\pi})\rangle\right|_{z^{+}=z^{i}_{\perp}=0};
\end{equation}
conversely, in a $\gamma^{\ast} \pi \to \gamma \rho$ process, only the vector quark TDA $V^{\pi\gamma}$ contributes. It  is defined as:
\begin{equation}\label{eq:VectorTDADef}
    \frac{i}{P^{+}}\frac{e}{f_{\pi}}\varepsilon_{\mu\nu\rho\sigma} n^{\mu}\epsilon^{\nu} P^{\rho} \Delta^{\sigma}_{\perp} V^{\pi\gamma}(x,\xi,u,\mu)=\frac{1}{2}\int\frac{dz^{-}}{2\pi}e^{ix P^{+}z^{-}}\left.\langle\gamma(q',\epsilon) |\bar{\psi}_{d}(-z/2)\gamma^{+}\psi_{u}(z/2)|\pi(p_{\pi})\rangle\right|_{z^{+}=z^{i}_{\perp}=0}
\end{equation}
where, as usual, $P=(p_{\pi}+q')/2$, $\Delta=q'-p_{\pi}$, $\varepsilon_{\mu\nu\rho\sigma}$ is the totally antisymmetric symbol, $f_{\pi}$ is the pion decay-constant and $\epsilon^{\mu}$ is polarization four-vector of the photon. The lightlike four-vector $n$ is such that, together with another lightlike vector $\tilde{n}$ ($n\cdot\tilde{n}=1$), any $v^{\mu}$ can be decomposed as $v^{\mu}=v^{+}\tilde{n}^{\mu}+v^{-}n^{\mu}+v^{\mu}_{\perp}$. Thus, $\pi - \gamma$ TDAs can be parametrized through the momentum fraction variable $x$; the skewness $\xi=\Delta^{+}/2P^{+}$, the standard Mandelstam variable $u=\Delta^{2}$ and a renormalization scale, $\mu$.
Transversity (chiral-odd) $\pi - \gamma$ TDAs, from their part, do not contribute to the processes we are discussing here, $\gamma^{\ast}\pi^{+}\rightarrow \gamma M^{+}$, for the same reason that chiral-odd quark GPDs in the nucleon do not contribute to leading twist DVCS, timelike Compton scattering or deeply virtual meson electroproduction \cite{Diehl:1998pd, Collins:1999un}. Thus we do not discuss them here.

Notably $\pi - \gamma$ TDAs fulfill a number of properties that follow directly from their field-theory definition. A comprehensive listing of those is out of the scope of this work (for details see \textit{e.g.} \cite{Pire:2004ie,Lansberg:2006fv} and references therein) but some of them shall be of interest to our presentation:
\begin{itemize}
    \item Support: $\pi - \gamma$ TDAs have support on the region $(x,\xi)\in[-1,1]\otimes [-1,1]$.
    \item Partonic interpretation: similarly to GPDs, an interpretation of $\pi - \gamma$ TDAs as parton-hadron scattering amplitudes can be constructed \cite{Pire:2004ie}. Accordingly, two kinematic regions can be distinguished: the DGLAP, $|x|>|\xi|$, and ERBL, $|x|<|\xi|$, domains.
    \item Polynomiality: following directly from their behavior under Lorentz transformations, a polynomiality property can be derived for $\pi - \gamma$ TDAs, guaranteeing that their Mellin moments behave as polynomials of a given degree in the skewness variable. Notice that, contrary to GPDs, TDAs are not invariant under time-reversal transformations and therefore these polynomials are neither even nor odd in $\xi$. As a consequence, they can be written in terms of Double Distributions \cite{Mezrag:2022pqk}.
    \item Sum rules: because the $z\rightarrow 0$ limits of Eq.~\eqref{eq:AxialTDADef} and Eq.~\eqref{eq:VectorTDADef} arise in weak radiative $\pi^{+}\rightarrow l^{+}\nu_{l}\gamma$ decays, we can relate them to axial and vector form factors, respectively \cite{Lansberg:2006fv}
    \begin{equation}\label{eq:FFsDef}
        \begin{array}{rcl}
            \displaystyle \int_{-1}^1 dx  A^{\pi\gamma} (x,\xi,u,\mu) & \displaystyle = & \displaystyle \frac{f_\pi}{m_{\pi}} F_{A} (u) \,,  \\
            \\
            \displaystyle \int_{-1}^1 dx  V^{\pi\gamma} (x,\xi,u,\mu)  & \displaystyle = & \displaystyle \frac{f_\pi}{m_{\pi}} F_{V} (u) \,,
        \end{array}
    \end{equation}
    imposing constraints on possible parametrizations of $\pi - \gamma$ TDAs.
\end{itemize}

So far, various models for $\pi - \gamma$ TDAs have been constructed on the basis of different frameworks \cite{Tiburzi:2005nj,Lansberg:2006fv,Broniowski:2007fs,Courtoy:2010qn,Zhang:2024dhs}. In this section, we shall present our own, constructed upon the overlap of lightfront wave functions (LFWFs) (see \cite{Diehl:2000xz}), a method primarily developed for GPDs. To do so we need a model for pion and photon LFWFs. From that point on, the overlap method allows to construct the DGLAP region of TDAs, which can be extended afterwards to the ERBL domain, providing us with a parametrization for axial and vector $\pi - \gamma$ TDAs.

\subsection{Lightfront wave functions}\label{subsec:LFWFs}


\subsubsection{LFWFs of the pion}\label{subsec:PionLFWFs}

The starting point for our calculation is the lowest Fock-space state contribution to the $\pi^+$ meson wave function \cite{Burkardt:2002uc}:
\begin{eqnarray}
    \label{eq:PionKet0}
    \ket{\pi^+,\uparrow \downarrow} &=& \int \frac{\textrm{d}k_\perp}{16\pi^3} \frac{\textrm{d}x \psi^{\pi}_{\uparrow \downarrow}(x,k_\perp)}{\sqrt{x(1-x)}} [ b_{u,\uparrow}^\dagger(x,k_\perp) d_{d,\downarrow}^\dagger(1-x,-k_\perp) - b_{u,\downarrow}^\dagger(x,k_\perp) d_{d,\uparrow}^\dagger(1-x,-k_\perp) ] \ket{0}\\
    \label{eq:PionKet1}
    \ket{\pi^+,\uparrow \uparrow} &=& \int \frac{\textrm{d}k_\perp}{16\pi^3} \frac{\textrm{d}x \psi^{\pi}_{\uparrow \uparrow}(x,k_\perp)}{\sqrt{x(1-x)}} [ (k_1 - ik_2) b_{u,\uparrow}^\dagger(x,k_\perp) d_{d,\uparrow}^\dagger(1-x,-k_\perp) \nonumber \\
    && + (k_1 + ik_2) b_{u,\downarrow}^\dagger(x,k_\perp) d_{d,\downarrow}^\dagger(1-x,-k_\perp) ] \ket{0},
\end{eqnarray}
where $\mathsf{P}=\uparrow\downarrow,\uparrow\uparrow$ label contributions from anti-aligned and aligned quark-helicity projections. The LFWFs \cite{Mezrag:2016hnp,Chouika:2017rzs}
\begin{eqnarray}
    \label{eq:PionLFWF0}
    \psi^\pi_{\uparrow\downarrow}(x,k_\perp) 
    &=& \frac{8 \sqrt{15} \pi}{\sqrt{N_c}} \frac{M^3}{(k_\perp^2 + M^2)^2} x(1-x), \\
    \label{eq:PionLFWF1}
    i k_\perp \psi^\pi_{\uparrow\uparrow}(x,k_\perp) & =& \frac{8 \sqrt{15} \pi}{\sqrt{N_c}}\frac{k_{\perp} M^{2}}{(k_{\perp}^{2}+M^{2})^{2}} x(1-x),
\end{eqnarray}
follow from a projection of the Bethe-Salpeter wave functions of the pion. Here $M$ is a mass scale fitted to $M=318~ \textrm{MeV}$ on experimental data for the $\pi^{+}$ electromagnetic form factor.


\subsubsection{LFWFs of the photon}\label{subsec:PhotonLFWFs}

The LFWFs of the photon can be derived based on the methods outlined in \cite{Dorokhov:2006qm,Ball:2002ps} (note that other definitions have been introduced \cite{Shi:2023jyk}). This time the relevant photon Fock-space states are:
\begin{eqnarray}
    \label{eq:PhotonUnpolarisedKet1}
    &&\ket{\gamma}_{\gamma \cdot n } = - e_q \int \frac{\textrm{d}^2\tilde{k}_\perp}{(2\pi)^3} \frac{\textrm{d}x}{2\sqrt{x(1-x)}}\epsilon \cdot k_\perp  \mathcal{N}_{\gamma \cdot n} \psi^{\gamma}_{\gamma \cdot n} (x,k_\perp)\nonumber \\
    &&\times\left( b^\dagger_{u,\uparrow}(x p_\gamma^+,k_\perp) d^\dagger_{u,\downarrow}((1-x)p_\gamma^+, -k_\perp ) + b^\dagger_{u,\downarrow}(x p_\gamma^+,k_\perp) d^\dagger_{u,\uparrow}((1-x)p_\gamma^+, -k_\perp )\right) \ket{0}\,,
    \\
    &&\ket{\gamma}_{\gamma \cdot n \gamma_5} =  - e_q\int  \frac{\textrm{d}^2k_\perp}{(2\pi)^3} \frac{\textrm{d}x}{2\sqrt{x(1-x)}} \mathcal{N}_{\gamma \cdot n \gamma_5} \psi^{\gamma}_{\gamma \cdot n \gamma_5} (x,k_\perp) i \varepsilon^{ij}k_i\epsilon_j\nonumber \\
    &&\times\left( b^\dagger_{u,\uparrow}(x p_\gamma^+,k_\perp) d^\dagger_{u,\downarrow}((1-x)p_\gamma^+, -k_\perp ) -  b^\dagger_{u,\downarrow}(x p_\gamma^+,k_\perp) d^\dagger_{u,\uparrow}((1-x)p_\gamma^+, -k_\perp )\right) \ket{0}\label{eq:photostatengng5},
    \\&&
        \ket{\gamma}_{\sigma^{n \perp}} = e_q\int  \frac{\textrm{d}^2k_\perp}{(2\pi)^3} \frac{\textrm{d}x}{2\sqrt{x(1-x)}} \mathcal{N}_{\sigma^{n \perp}} \psi^\gamma_{\sigma^{n \perp}} (x,k_\perp) \Big((\epsilon^1 -i\epsilon^2) b^\dagger_{u,\uparrow}(x p_\gamma^+,k_\perp) d^\dagger_{u,\uparrow}((1-x)p_\gamma^+, -k_\perp ) \nonumber \\
    && - (\epsilon^1 + i\epsilon^2) b^\dagger_{u,\downarrow}(x p_\gamma^+,k_\perp) d^\dagger_{u,\downarrow}((1-x)p_\gamma^+, -k_\perp )\Big) \ket{0}.
\end{eqnarray}
where $\Gamma$ in $\ket{\gamma}_\Gamma$ indicates the Dirac projector selecting the associated LFWFs $\psi^{\gamma}_\Gamma$. Again, following \cite{Ball:2002ps} and our previous efforts in the pion case \cite{Mezrag:2016hnp,Chouika:2017rzs,Chavez:2021llq} we find
\begin{equation}\label{photonlfwfs1}
    \begin{array}{rcr}
    \displaystyle \psi^{\gamma}_{\gamma \cdot n}(x,k_\perp) & \displaystyle =& \displaystyle   
    -\frac{1-2x}{k^{2}_{\perp}+M^{2}},\\
    \\
    \displaystyle\psi^{\gamma}_{\gamma\cdot n\gamma_5}(x,k_\perp) & \displaystyle =& \displaystyle \frac{1}{k^{2}_{\perp}+M^{2}},\\
    \\
    \displaystyle\psi^{\gamma}_{\sigma^{n_\perp}}(x,k_\perp) & \displaystyle = & \displaystyle \frac{iM}{k^{2}_{\perp}+M^{2}}.
    \end{array}
\end{equation}

 \begin{figure}
    \centering
     \includegraphics[width=0.6\textwidth]{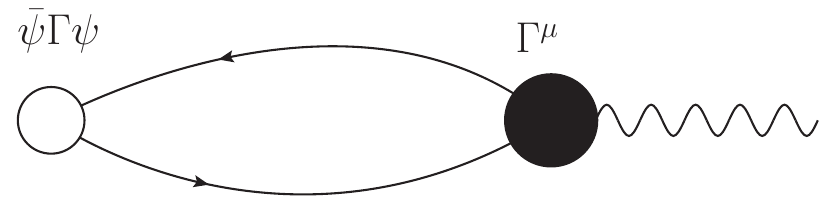}
    \caption{Graphical representation of the formulae used to compute the LFWFs defined by the operator $\bar{\psi}\Gamma \psi$ from the quark-photon vertex $\Gamma^{\mu}$.}
     \label{fig:Photon_LFWFs_computation}
 \end{figure}

Albeit very refined ingredients for the quark propagator (see for instance \cite{Aguilar:2018epe,Gao:2021wun}) or the quark photon vertex exist \cite{Qin:2013mta,Binosi:2016rxz,Mezrag:2020iuo} allowing for instance to express a large number of the vertex form factors as combinations of the quark self energy ones, we stick in this paper to a simple tree-like vertex, in order to simply assess the order of magnitude of backward Sullivan DVCS.
Note however that, when combining the Longitudinal and Transverse Ward-Takahashi identities for the photon vertex \cite{Ball:1980ay,Qin:2013mta}, most of its form factors can be derived from the quark self-energy. 
Similarly, because of the quark-level Goldberger-Treiman relation \cite{Maris:1997hd}, the pion Bethe-Salpeter wave function is also expressed in terms of the quark self energy in the chiral limit.
This in principle, allows one to connect the pion-to-photon TDA with a mechanism of quark mass generation, expressed in the quark two-point function. 
However, as highlighted in \cite{Mezrag:2020iuo}, this task is rather involved and is left for future work.
The computation of the LFWFs is then performed following the Mellin Moment technique developed for Nakanishi integrals \cite{Mezrag:2014tva,Mezrag:2014jka,Chouika:2017rzs,Mezrag:2017znp,Mezrag:2018hkk} and is sketched in Fig.~\ref{fig:Photon_LFWFs_computation}. Finally, we fix the overall phase so that $\psi^{\gamma}_{\gamma\cdot n}$ and $\psi^{\gamma}_{\gamma\cdot n\gamma_5}$ are consistent with the results found in \cite{Ball:2002ps}. We noted however a phase difference in $\psi_{\sigma^{n_\perp}}$, but such a phase shift compensates with the pion one to give a result in agreement with the data regarding the constraints \eqref{eq:FFsDef}.
We leave the investigation of such a shift for future work, as our primary target here is to provide a rough estimate of the cross-section. Eqs.~\eqref{eq:PhotonUnpolarisedKet1} to \eqref{photonlfwfs1} should thus be seen as the basis of our modelling strategy for the photon.

\subsubsection{The anomalous photon GPD}

An interesting consistency check of our approach is the calculation of the anomalous photon GPDs from the overlap of the point-like photon LFWFs. Let us briefly remind the reader that the deep inelastic structure functions on a photon target have their shapes and magnitudes determined by a short distance analysis in QCD \cite{Witten:1977ju}. The resulting quark (and gluon) PDFs are proportional to $\log Q^2$ and are called anomalous. In a similar way, there are  two chiral-even anomalous leading twist photon GPDs \cite{Friot:2006mm,Gabdrakhmanov:2012aa}, a vector and an axial one, and their $(x,\xi)$ behaviours have been determined (at $t=0$) both in the DGLAP and ERBL regions.
Within the overlap method, the photon to photon vector GPD expression for the DGLAP region is, after integration on $d^2k_\perp$ and considering only the logarithmic contribution,
\begin{eqnarray}
    H^q(x,\xi,0) &=&  \frac{e^2_q N_c}{4\pi^2} \frac{(x^2 + (1-x)^2 - \xi^2)}{1 - \xi^2} \log \left(\frac{Q^2}{M^2}\right) \Theta(x - \xi)\,,
\end{eqnarray}
in agreement with the result of the leading order perturbative calculation \cite{Friot:2006mm}.

\subsection{DGLAP region: TDAs in the overlap representation}\label{subsec:DGLAPOverlap}

Once we have introduced our models for the pion and photon LFWFs we are in a position to face the extraction of $\pi - \gamma$ TDAs. As it was already said, to this end we exploit the overlap representation \cite{Diehl:2000xz}. Formally there is an infinite number of those LFWFs contributing to the representation. This is because there are infinitely many states --with an arbitrary number of partons-- contributing to the Fock-space decomposition of the relevant states, here pion- and photon-states. A practical calculation requires a truncation of the relevant Fock-space expansions. In practice, this step introduces an important limitation of the method: only the DGLAP region can be accessed. To understand why one needs to recall that, in the DGLAP region, only Fock-space states containing the same number of partons overlap. This feature can be read from the partonic interpretation of TDAs. Meanwhile, in the ERBL region, states with different number partons are connected. Accordingly, there is no truncation of the Fock-space expansions that allows to access, in a consistent manner, both kinematic regions.

In this work we stick to a valence approximation, assuming the parton content of the external pion- and photon-states to be well approximated by a two-body truncation of the corresponding Fock-space expansions. The  pion and photon LFWF models presented in Secs.~\ref{subsec:PionLFWFs} and \ref{subsec:PhotonLFWFs} then allow us construct the desired model for axial and vector $\pi - \gamma$ TDAs within the DGLAP region. 
Following these steps we find vector and axial $\pi - \gamma$ TDAs to decompose into contributions of definite quark (in the pion) helicity projections\footnote{For shortness in the notation, in the following we shall drop any explicit labeling of the renormalization scale $\mu$. We will restore it in Sec.~\ref{subsec:ScaleEvolution} when we will turn to a discussion of scale-evolution effects.},
\begin{equation}\label{eq:HelicityDecomposition}
    \begin{array}{rcl}
    \displaystyle A^{\pi\gamma}(x,\xi,u) & \displaystyle = & \displaystyle A^{\pi\gamma}_{\uparrow\downarrow}(x,\xi,u)+A^{\pi\gamma}_{\uparrow\uparrow}(x,\xi,u),\\
    \\
    \displaystyle V^{\pi\gamma}(x,\xi,u) & \displaystyle = & \displaystyle V^{\pi\gamma}_{\uparrow\downarrow}(x,\xi,u)+V^{\pi\gamma}_{\uparrow\uparrow}(x,\xi,u).
    \end{array}
\end{equation}
Moreover, within the DGLAP region\footnote{For definiteness we quote here, and in the following, results for $\xi>0$.} each of those such pieces receive contributions from two kinematic regions: the ``quark part'' of the DGLAP domain, $x\geq\xi$, and the ``antiquark'' one, $x\leq -\xi$
\begin{equation}
\begin{array}{rcl}\label{eq:HelicityDecompositionDGLAP}
    \displaystyle\left.A^{\pi\gamma}_{\mathsf{P}}(x,\xi,u)\right|_{|x|>\xi} & \displaystyle = & \displaystyle e_{d}\theta(x\geq\xi)a^{\textrm{DGLAP}}_{\mathsf{P}}(x,\xi,u)+(-1)^{L}e_{u}\theta(x\leq -\xi)a^{\textrm{DGLAP}}_{\mathsf{P}}(-x,\xi,u),\\
    \\
    \displaystyle\left.V^{\pi\gamma}_{\mathsf{P}}(x,\xi,u)\right|_{|x|>\xi} & \displaystyle = & \displaystyle e_{d}\theta(x\geq\xi)v^{\textrm{DGLAP}}_{\mathsf{P}}(x,\xi,u)-e_{u}\theta(x\leq -\xi)v^{\textrm{DGLAP}}_{\mathsf{P}}(-x,\xi,u),
\end{array}
\end{equation}
where $L=0,1$ for $\mathsf{P}=\uparrow\downarrow,\uparrow\uparrow$, respectively; and where we have introduced the short-hand notation
\begin{equation}
    \theta(x\geq\xi)=\theta\left(\frac{x-\xi}{1-\xi}\right)\theta\left(\frac{x+\xi}{1+\xi}\right),\qquad\theta(x\leq -\xi)=\theta\left(\frac{x+\xi}{\xi-1}\right)\theta\left(\frac{\xi-x}{1+\xi}\right).
\end{equation}

Using the overlap method and the two-body LFWFs discussed before we have found
\begin{equation}\label{eq:ModelPieces}
\begin{array}{rcl}
\displaystyle a^{\textrm{DGLAP}}_{\uparrow\downarrow}(x,\xi,u) & \displaystyle = &\displaystyle \mathcal{N}\frac{(1-x)^{2}(x+\xi)}{(1-\xi^{2})^{2}(1+\xi)}\frac{(1+\xi-2x)}{\zeta(\zeta+1)}\left[1+2\zeta-\frac{1}{\zeta+1}\frac{\tanh^{-1}\left(\sqrt{\frac{\zeta}{\zeta+1}}\right)}{\sqrt{\frac{\zeta}{\zeta+1}}}\right],\\
\\
\displaystyle a^{\textrm{DGLAP}}_{\uparrow\uparrow}(x,\xi,u) &\displaystyle = &\displaystyle \mathcal{N}\frac{(1-x)^{2}(x+\xi)}{(1-\xi^{2})^{2}(1+\xi)}\frac{1-\xi}{\zeta(\zeta+1)}\left[-1+\frac{(2\zeta+1)}{\zeta+1}\frac{\tanh^{-1}\left(\sqrt{\frac{\zeta}{\zeta+1}}\right)}{\sqrt{\frac{\zeta}{\zeta+1}}}\right],
\end{array}
\end{equation}
for the axial distribution, where $\mathcal{N}$ is a normalization factor reading
\begin{equation}
\mathcal{N}=\frac{\sqrt{15 N_{c}}}{8\pi M}f_{\pi},
\end{equation}
and $\zeta$ is a parameter controlling the $u$-dependence of the TDA
\begin{equation}\label{eq:udependence}
\zeta=-\frac{u}{4 M^{2}}\frac{(1-x)^{2}}{(1-\xi^{2})},~\qquad\textrm{with }u\leq 0.
\end{equation}

Similarly, for the vector TDA we obtained
\begin{equation}
\begin{array}{rcl}
\displaystyle v^{\textrm{DGLAP}}_{\uparrow\downarrow}(x,\xi,u,\mu)& \displaystyle = & \displaystyle \mathcal{N}\frac{(1-x)^{2}(x+\xi)}{(1-\xi^{2})^{2}(1+\xi)}\frac{(1-\xi)}{\zeta(\zeta+1)}\left[1+2\zeta-\frac{1}{\zeta+1}\frac{\tanh^{-1}\left(\sqrt{\frac{\zeta}{\zeta+1}}\right)}{\sqrt{\frac{\zeta}{1+\zeta}}}\right],\\
\\
\displaystyle v^{\textrm{DGLAP}}_{\uparrow\uparrow}(x,\xi,u,\mu) & \displaystyle = & \displaystyle \mathcal{N}\frac{(1-x)^{2}(x+\xi)}{(1-\xi^{2})^{2}(1+\xi)}\frac{(1-\xi)}{\zeta(\zeta+1)}\left[-1+\frac{(2\zeta+1)}{\zeta+1}\frac{\tanh^{-1}\left(\sqrt{\frac{\zeta}{\zeta+1}}\right)}{\sqrt{\frac{\zeta}{\zeta+1}}}\right].
\end{array}
\end{equation}

Notice that, as expected, the model shows an entangled $x,\xi,u$ dependence. Moreover, that the resulting TDA models fulfill the support property of TDAs can be seen already at this point. This is a direct consequence of the support property of LFWFs.

\subsection{Transition form factors}

\begin{figure}[t]
 	\begin{subfigure}[b]{0.45\textwidth}
        \centering
        \includegraphics[scale=0.85]{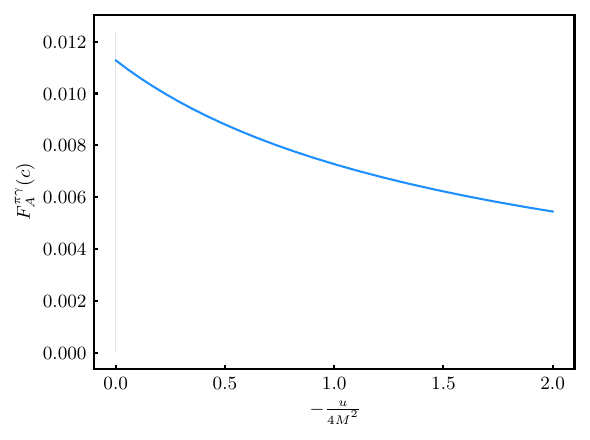}
    \caption{Axial transition form factor}
    \end{subfigure} \hfill
    \begin{subfigure}[b]{0.45\textwidth}
        \centering
        \includegraphics[scale=0.85]{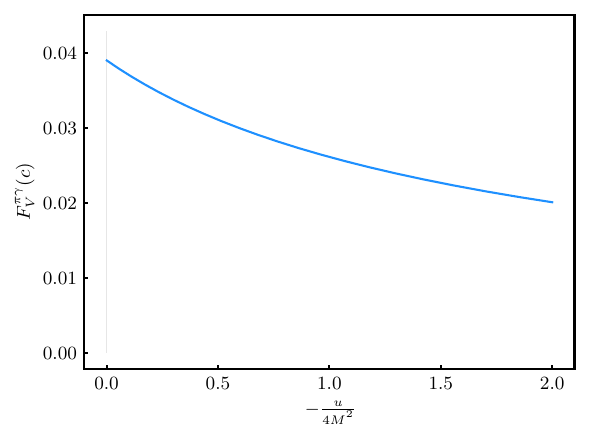}
    \caption{Vector transition form factor}
    \end{subfigure}
    \caption{\label{fig:tFFs} Axial and vector $\pi-\gamma$ transition form factors computed from our two-body models for the axial and vector TDAs. The variable $c$ is defined as $c=-u/4M^{2}$.}
\end{figure}

Once we have built our models for the axial and vector TDAs, we can exploit them for the evaluation of the $\pi-\gamma$ transition form factors, Eqs.~\eqref{eq:FFsDef}, benchmarking them against sum rules. Indeed, as entailed by Lorentz covariance, transition form factors do not depend on the skewness variable $\xi$. Accordingly, one can extract them from the knowledge of DGLAP TDAs, only. Furthermore, one can safely compute them using $\xi\rightarrow 0$ expression for the corresponding TDAs and at any $\mu$. We proceed in that way, which allows to perform the calculation in closed form. The results are shown in Fig.~\ref{fig:tFFs}.

What is more interesting though it to concentrate on the limit $u=0$, where we recover the axial and vector transition form factors measured in weak decays $\pi^{+}\rightarrow l^{+}\nu_{l}\gamma$. In those cases we obtain
\begin{equation}
    \begin{array}{rcl}
        F_{A}^{\pi\gamma}(0) & \displaystyle = &\displaystyle 0.0113,\\
        \\
        F_{V}^{\pi\gamma}(0) & \displaystyle = &\displaystyle 0.039,
    \end{array}
\end{equation}
where we used $f_{\pi}=130.02\pm 1.2~\textrm{MeV}$ and $m_{\pi}=139.57039\pm 0.00018$ \cite{PDG:2024cfk}. Our results can be compared with reported experimental values
\begin{equation}
    \label{eq:PDG2024FormFactors}
    \begin{array}{rcl}
        \left.F_{A}^{\pi\gamma}(0)\right|_{\textrm{PDG2024}~}
        & \displaystyle = &\displaystyle 0.0119\pm 0.0001,\\
        \\
        \left.F_{V}^{\pi\gamma}(0)\right|_{\textrm{PDG2024}~}
         & \displaystyle = &\displaystyle 0.0254\pm 0.0017.
    \end{array}
\end{equation}

For the axial case, the $5\%$ discrepancy is surprisingly good, being given the simplicity of the models we used. However, the vector part is off by roughly 50\% . This disagreement is similar to what has been obtained up to now in the literature. Using TDA calculations in a Nambu--Jona-Lasinio model, results for the axial and vector transition form factors are reported, independently, in \cite{Courtoy:2008ij,Zhang:2024dhs}, and using a quark-model in \cite{Broniowski:2007fs}. The results of all those works for the vector transition form factors are consistent with ours.

\subsection{ERBL region: The covariant extension}

\begin{figure}[t]
\centering
    \begin{subfigure}[H]{0.45\textwidth}
        \centering
        \includegraphics[scale=0.8]{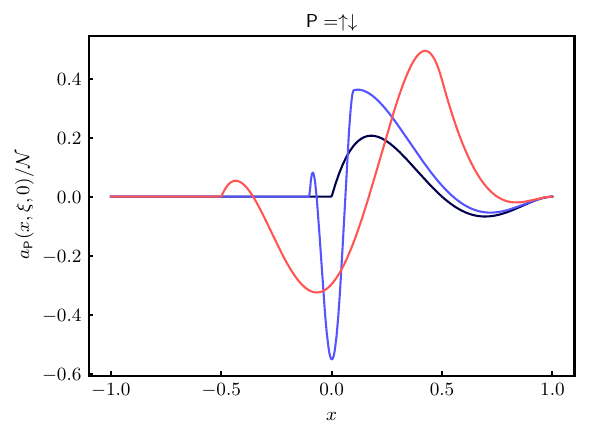}
    \end{subfigure} \hfill
    \begin{subfigure}[H]{0.45\textwidth}
        \centering
        \includegraphics[scale=0.8]{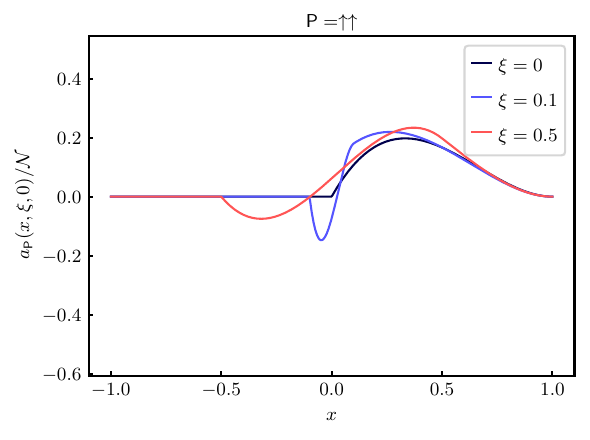}
    \end{subfigure} \\
    \vspace{1cm}
    \begin{subfigure}[H]{0.45\textwidth}
        \centering
        \includegraphics[scale=0.8]{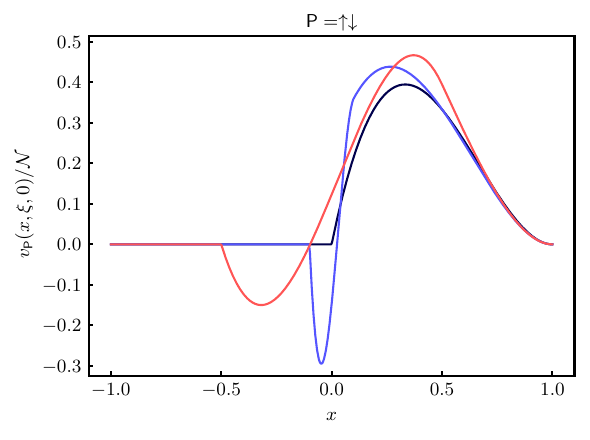}
    \end{subfigure} \hfill
    \begin{subfigure}[H]{0.45\textwidth}
        \centering
        \includegraphics[scale=0.8]{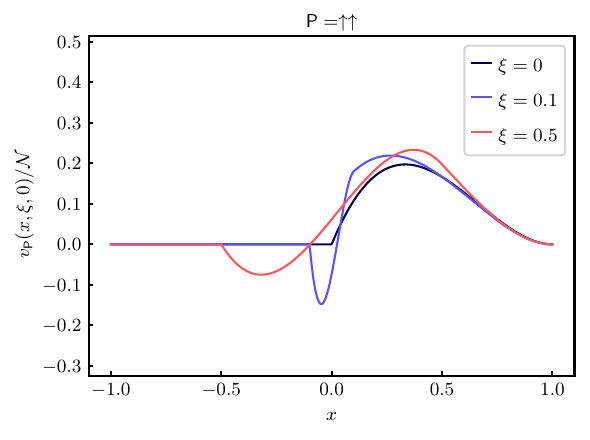}
    \end{subfigure}
\caption{Illustration of the covariant extension strategy as applied to the axial (\textsc{Top panel}) and vector (\textsc{Bottom panel}) TDAs, in the $u\rightarrow 0$ and selected values of the skewness variable, $\xi=0,0.1,0.5$ (dark-blue, light-blue and red, respectively). The left panels show the contributions for quarks-in-pion with anti-aligned helicity projection while the right-hand side plots display the corresponding results when the quarks have aligned helicity projections.}
\label{fig:CovExtu0}
\end{figure}

As explained in Sec.~\ref{subsec:DGLAPOverlap}, a direct access to the ERBL region through the overlap representation is precluded. However, in order to exploit our TDA models for the description of scattering processes both regions must be known. To overcome this difficulty we employ the covariant extension strategy \cite{Chouika:2017dhe, Chavez:2021llq, DallOlio:2024vjv} which guarantees that, given a TDA within the DGLAP domain, its ERBL region is uniquely\footnote{Importantly, TDAs being flavor non-singlet objects, no D-term--like ambiguity arises, in contrast to the GPD case.} determined. In a nutshell: as GPDs, $\pi - \gamma$ TDAs benefit from a representation as the Radon transform of double distributions \cite{Teryaev:2001qm}. Provided that the solution to the inverse Radon transform problem exists and is unique, when TDAs are known only on the DGLAP region \cite{Chouika:2017dhe}, the associated double distributions can be found and employed afterwards to reconstruct the ERBL domain \cite{Chavez:2021llq, DallOlio:2024vjv}. 

\subsubsection{\texorpdfstring{Covariant extension in the $u\rightarrow 0$ limit}{Covariant extension in the u->0 limit}}

For illustration, let us first consider the simpler $u=0$ case. In this kinematic limit, for the model constructed in the previous section, a solution to the inverse Radon transform problem can be found in analytic form. Indeed, we may find the associated double distributions (in Polyakov-Weiss scheme) to be polynomials in the  kinematic variables $\beta$ and $\alpha$:
\begin{equation}\label{eq:ddt0}
\begin{array}{rcl}
\displaystyle h^{A}_{\uparrow\downarrow}\left(\beta,\alpha,0\right)/\mathcal{N} & \displaystyle = & \displaystyle -\frac{1}{3}+\frac{10}{3}\alpha+\alpha^{2}-4\alpha^{3}+\frac{10}{3}\beta-6\alpha\beta-4\alpha^{2}\beta-7\beta^{2}+4\alpha\beta^{2}+4\beta^{3},\\
\\
\displaystyle h^{A}_{\uparrow\uparrow}\left(\beta,\alpha,0\right)/\mathcal{N} & \displaystyle = & \displaystyle \frac{1}{6}+\frac{\alpha}{3}-\frac{\alpha^{2}}{2}+\frac{\beta}{3}-\alpha\beta-\frac{\beta^{2}}{2},
\end{array}
\end{equation}
for the axial case, and
\begin{equation}
\begin{array}{rcl}
\displaystyle h^{V}_{\uparrow\downarrow}(\beta,\alpha,0)/\mathcal{N} & \displaystyle = & \displaystyle \frac{1}{3}+\frac{2}{3}\alpha-\alpha^{2}+\frac{2}{3}\beta-2\alpha\beta-\beta^{2},\\
\\
\displaystyle h^{V}_{\uparrow\uparrow}(\beta,\alpha,0)/\mathcal{N} & \displaystyle = & \displaystyle \frac{1}{6}+\frac{\alpha}{3}-\frac{\alpha^{2}}{2}+\frac{\beta}{3}-\alpha\beta-\frac{\beta^{2}}{2},
\end{array}
\end{equation}
for the vector case; such that
\begin{equation}\label{eq:RTav}
    \begin{array}{rcl}
        \displaystyle a_{\mathsf{P}}(x,\xi,0) & \displaystyle = & \displaystyle \int_{\Omega} d\beta d\alpha\delta(x-\beta-\alpha\xi)h^{A}_{\mathsf{P}}(\beta,\alpha,0),\\
        \\
        \displaystyle v_{\mathsf{P}}(x,\xi,0) & \displaystyle = & \displaystyle \int_{\Omega} d\beta d\alpha\delta(x-\beta-\alpha\xi)h^{V}_{\mathsf{P}}(\beta,\alpha,0).
    \end{array}
\end{equation}

Notice that, in contrast to the GPD case, the double distributions associated to TDAs are not even in the variable $\alpha$. This is a direct consequence of the lost of time-reversal invariance, as it is the fact that Mellin moments of TDAs are (nor even, nor odd) polynomials in the skewness variable, alike those of GPDs.

Given the above double distributions, our TDAs can be straightforwardly extended to their ERBL domain through a direct evaluation of the Radon transform. The results, Fig.~\ref{fig:CovExtu0}, read:
\begin{equation}
\begin{array}{rcl}
\displaystyle \left. a^{\textrm{ERBL}}_{\uparrow\downarrow}\left(x,\xi,0\right)\right|_{|x|\leq\xi}&\displaystyle = & \displaystyle \frac{\mathcal{N}}{3\xi^{4}\left(1+\xi\right)^{3}}\left[-2\xi^{4}\left(1+\xi\right)+x^{3}\xi\left(1-13\xi^{2}\right)+x^{2}\xi^{2}\left(5+\xi\left(20+3\xi\right)\right)-\right.\\
&&\displaystyle \\
&&\hfill\left.-~x\xi^{3}\left(1-\xi\left(8+5\xi\right)\right)-x^{4}\left(3+\xi\left(10+11\xi\right)\right)\right],\\
\\
\displaystyle \left. a^{\textrm{ERBL}}_{\uparrow\uparrow}\left(x,\xi,0\right)\right|_{|x|\leq\xi} & \displaystyle = & \displaystyle -\frac{\mathcal{N}\left(x+\xi\right)}{6\xi^{3}\left(1+\xi\right)^{3}}\left[\xi^{2}-\xi^{3}\left(4+\xi\right)-2x\xi\left(1+\left(4-\xi\right)\xi\right)+x^{2}\left(1+\xi\left(4+7\xi\right)\right)\right].
\end{array}
\end{equation}
for the axial TDA, and
\begin{equation}
\begin{array}{rcl}
\displaystyle \left.v^{\textrm{ERBL}}_{\uparrow\downarrow}(x,\xi,0)\right|_{x\leq |\xi|} & \displaystyle = & \displaystyle -\frac{\mathcal{N}(x+\xi)}{3\xi^{3}(1+\xi)^{3}}\left[\xi^{2}-\xi^{3}(4+\xi)-2x\xi(1+(4-\xi)\xi)+x^{2}(1+\xi(4+7\xi))\right]\\
\\
\displaystyle \left.v^{\textrm{ERBL}}_{\uparrow\uparrow}(x,\xi,0)\right|_{x\leq |\xi|} & \displaystyle = & \displaystyle -\frac{\mathcal{N}(x+\xi)}{6\xi^{3}(1+\xi)^{3}}\left[\xi^{2}-\xi^{3}(4+\xi)-2x\xi(1+(4-\xi)\xi)+x^{2}(1+\xi(4+7\xi))\right];\\
\end{array}
\end{equation}
for the vector case.

Notably, the results obtained and shown for selected values of skewness in Fig.~\ref{fig:CovExtu0} display the expected properties. First, the curves are continuous at the crossing-points $x=\pm\xi$. Moreover, that they exhibit the expected polynomiality property can be read directly from Eq.~\eqref{eq:RTav} or checked afterwards by direct integration.

\begin{figure}[t]
\centering
    \begin{subfigure}[H]{0.45\textwidth}
        \centering
        \includegraphics[scale=0.9]{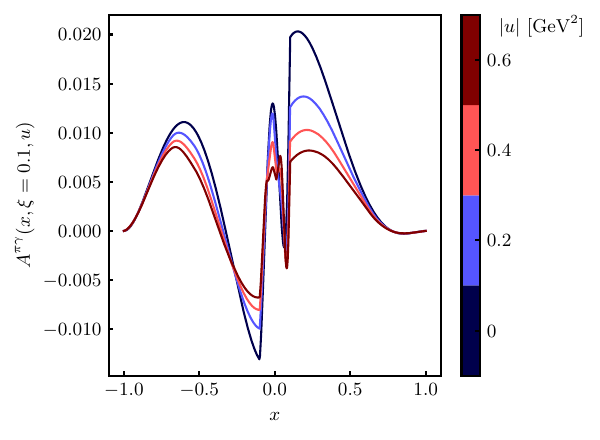}
    \end{subfigure} \hfill
    \begin{subfigure}[H]{0.45\textwidth}
        \centering
        \includegraphics[scale=0.9]{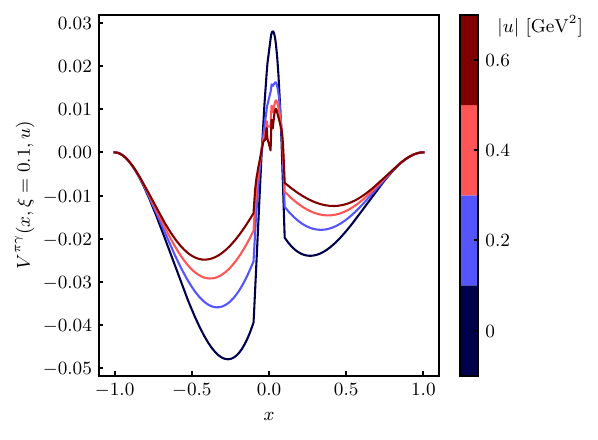}
    \end{subfigure} \\
    \vspace{1cm}
        \begin{subfigure}[H]{0.45\textwidth}
        \centering
        \includegraphics[scale=0.9]{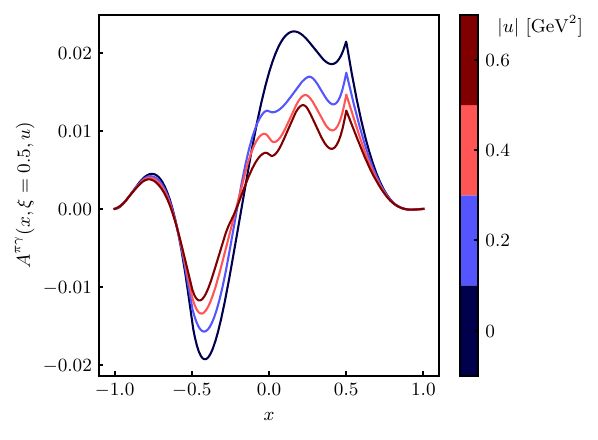}
    \end{subfigure} \hfill
    \begin{subfigure}[H]{0.45\textwidth}
        \centering
        \includegraphics[scale=0.9]{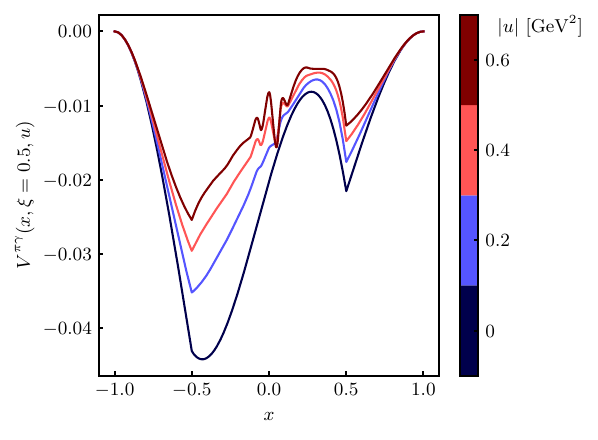}
    \end{subfigure}
\caption{\label{fig:CovExt} Illustration of the covariant extension strategy for two values of the skewness variable, $\xi=0.1,0.5$, shown in the \textsc{Top-} and \textsc{Bottom Panel}, respectively; and a number of values of the momentum transfer variable, $|u|=0,0.2,0.4,0.6~\textrm{GeV}^{2}$. Left-hand side plots show results for the axial TDA while right-hand side figures do so for the vector TDA.}
\end{figure}

\subsubsection{\texorpdfstring{Covariant extension in the general $u\neq 0$ case}{Covariant extension in the general u different from 0 case}}

For the general case where $u < 0$, the covariant extension strategy can be identically applied. The challenge in this case is that no simple analytic expression for the underlying double distributions can be envisaged. For this reason we resort to a numerical implementation of the method \cite{Chavez:2021llq,DallOlio:2024vjv} as available in \texttt{PARTONS} \cite{Berthou:2015oaw} which employs a discretization of the double distribution domain and an approximation of the corresponding structure through finite element methods and second-order Lagrange polynomials. For the axial TDA we setup our numerical procedure building a mesh with maximum element-area of $0.09$, which results into $n_{e}=14$ triangles. We sample such a domain with $12 n_{e}$ DGLAP lines. For the vector TDA we use $n_{e}=59$ elements to discretize the double distribution domain (maximum element-area of $0.025$) which are sampled with $30n_{e}$ lines.

As an illustration of the procedure Fig.~\ref{fig:CovExt} shows our results for the axial and vector TDAs at $\xi=0.1$ and $0.5$ and for $|u|=0,0.2,0.4,0.6~\textrm{GeV}^{2}$, covering the range on which one is often interested when studying the Sullivan process.

\subsection{Scale evolution}\label{subsec:ScaleEvolution}

\begin{figure}[htb]
\centering
    \begin{subfigure}[H]{0.45\textwidth}
        \centering
        \includegraphics[scale=0.8]{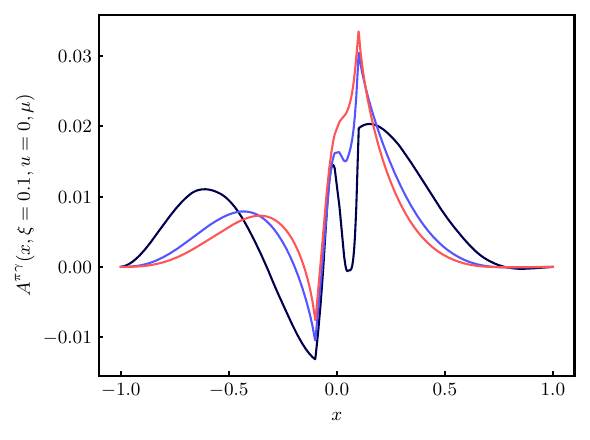}
    \end{subfigure} \hfill
    \begin{subfigure}[H]{0.45\textwidth}
        \centering
        \includegraphics[scale=0.8]{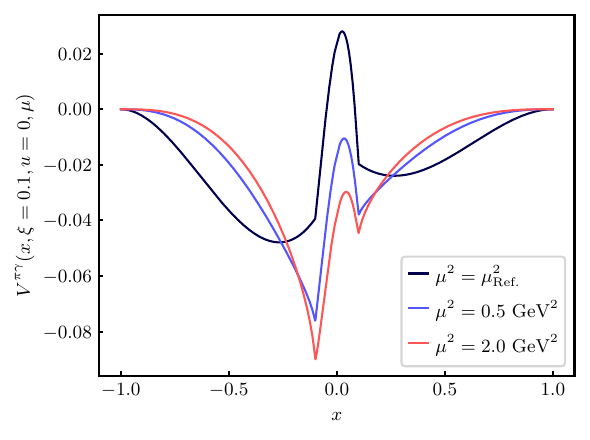}
    \end{subfigure} \\
    \vspace{1cm}
        \begin{subfigure}[H]{0.45\textwidth}
        \centering
        \includegraphics[scale=0.8]{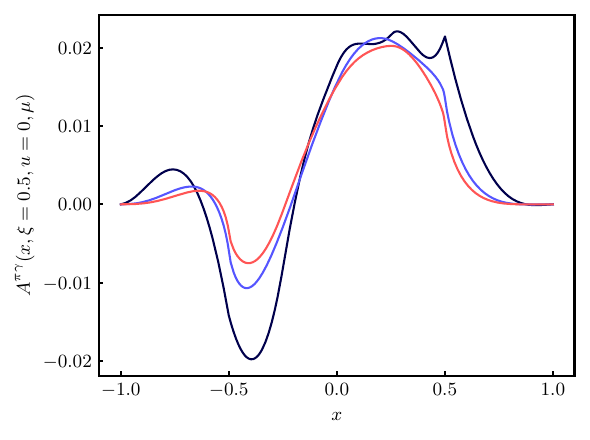}
    \end{subfigure} \hfill
    \begin{subfigure}[H]{0.45\textwidth}
        \centering
        \includegraphics[scale=0.8]{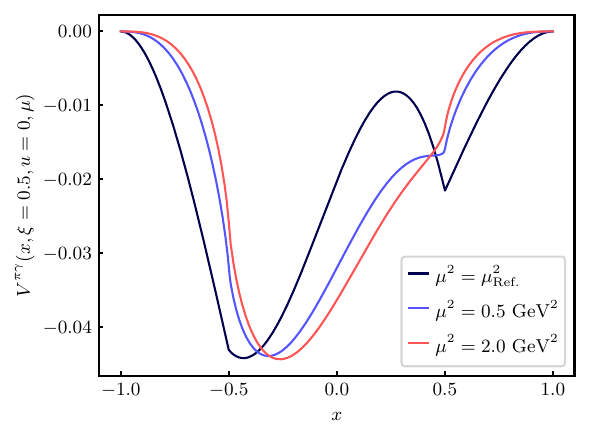}
    \end{subfigure}
\caption{\label{fig:Evolution} Illustration of the effect of scale evolution, implemented as described in Sec.~\ref{subsec:ScaleEvolution}, on our axial (\textsc{Left-}) and vector (\textsc{Right panel}) TDAs. Results are shown for $u=0$ (as TDA evolution equations do not depend on the momentum transfer variable) and two values of skewness variable, $\xi=0.1,0.5$, visible in the top-  and bottom-rows, respectively.}
\end{figure}

We want to estimate of the cross-section for $ep\rightarrow e\gamma M n$ at future facilities, in backward and Bjorken kinematics. To this end the necessary ingredient is a sensible parametrization of vector and axial $\pi-\gamma$ TDAs. After following the procedure described in the previous sections we are given with those such models which, in addition to fulfill the necessary properties, give acceptable estimations for the axial and vector the pion-photon transition form factors. However we are still a step back in our aim to describe backward DVCS: we need to control the scale evolution of our TDA models. Two important questions need to be addressed in this regard. First, how do TDAs evolve with scale? And second, what is the scale at which our models are defined?

The first question was already addressed in \cite{Pire:2004ie} showing that meson-to-photon TDAs evolve like non-singlet GPDs.
Regarding the second point, we argue that our TDA models are defined at some low renormalization scale $\mu_{\textrm{R}}=\mu_{\textrm{Ref.}}$, corresponding to a resolution scale $\lambda_{\textrm{Ref.}}$ which allows only to resolve the two valence degrees of freedom, in concordance with the two-body truncation of the Fock-space expansion we employed for the formulation of our model.

To fix the scale, we stick to the model for evolution developed elsewhere \cite{Cui:2022bxn} and already used in the case of forward Sullivan DVCS \cite{Chavez:2021llq,Chavez:2021koz}. In a nutshell, leading order evolution equations may be extended to the non-perturbative regime through the absorption of 
``non-perturbative corrections'' into a redefinition of the strong coupling as $\alpha_{s}^{\textrm{Eff.}}$ which, saturating in the infrared regime \cite{Binosi:2016nme,Cui:2019dwv,Pelaez:2021tpq}, approaches the standard $\overline{\textrm{MS}}$ coupling in the perturbative region. Among the possible choices for such an effective charge, the authors in \cite{Cui:2022bxn} suggest the use of that introduced in \cite{Cui:2019dwv,Cui:2020dlm} which, in combination with the above strategy, has proved to give reliable results in PDF and GPD studies of the pion \cite{Yin:2023dbw}. Within this setup, a value of the scale $\mu_{\textrm{Ref.}}$ can be assigned as that scale where long-range modes dominate over short-range interactions, preventing the effective coupling from blowing-up; say at the former position of the Landau pole. This argument gives, for the effective coupling \cite{Cui:2019dwv,Cui:2020dlm}, $\mu_{\textrm{Ref.}}=0.331~\textrm{GeV}$.

In line with the previous construction we claim that our TDA models are defined at such a scale $\mu_{\textrm{Ref.}}$ and that, using the model for evolution described in \cite{Cui:2022bxn} from that scale on, we can reliably translate them to energy scales relevant for collider experiments. The only task is to solve the corresponding evolution equations. To this end we take advantage of the \texttt{Apfel++} library \cite{Bertone:2017gds,Bertone:2013vaa} which implements leading order evolution equations for GPDs, and therefore for TDAs \cite{Bertone:2022frx,Bertone:2023jeh}. As an illustration we show results for evolution in Fig.~\ref{fig:Evolution}. Importantly, as an aside result of the convolution with the evolution kernels, the evolved TDAs are smoother than the ones at the reference scale. This is due to the integration over the entire $x$ range which, numerically, has the effect of filtering high-frequency oscillations. Notice also that the evolved distributions show the expected cusp at the transition points $x=\pm\xi$, which is maintained by the structure of the leading order evolution kernels. Finally, we would like to mention here that, in contrast to \textit{e.g.} the GPD case, TDAs being non-singlet objects, no gluon coupling exists at the level of evolution. Accordingly, we expect the present evolved results to provide a good first approximation of the TDAs which enter backward DVCS scattering amplitudes at future facilities.

\section{The backward DVCS through the Sullivan process}
\label{sec:Amplitude}

\begin{figure}[b]
\centering
    \scalebox{0.7}[0.7]
    {
    \begin{subfigure}[H]{0.45\textwidth}
        \centering
        \includegraphics[scale=1]{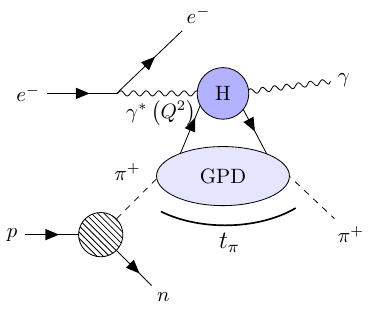}
    \end{subfigure}
    }
    \scalebox{0.7}[0.7]
    {
        \begin{subfigure}[H]{0.45\textwidth}
            \centering
            \includegraphics[scale=1]{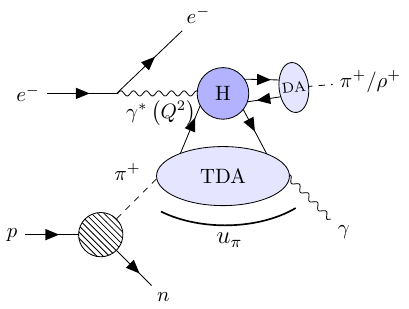}
        \end{subfigure}
    }
    \scalebox{0.7}[0.7]
    {
        \begin{subfigure}[H]{0.45\textwidth}
            \centering
            \includegraphics[scale=1]{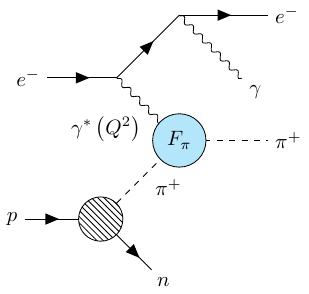}
        \end{subfigure}
    }
\caption{\textsc{Left panel}: forward DVCS; \textsc{Central panel}: backward DVCS; \textsc{Right panel}: the Bethe-Heitler process, in the Sullivan reaction  $e^-(l) + p(p)  \rightarrow  e^-(l') + \gamma(q') + M^+(p'_{M}) + n(p')$.  }
\label{fig:kin}
\end{figure}

We have described our modeling strategy for axial and vector $\pi-\gamma$ TDAs, including the effect of scale evolution. We are therefore in a position to exploit these models for the description of $ep\rightarrow e\gamma Mn$ scattering. Indeed, when the momentum transfer between the initial- and final-state nucleons remains small, say near the threshold for pion production, the same scattering process can be seen as being mediated by the scattering of the electron on a pion emitted by the nucleon, Fig.~\ref{fig:kin}; \textit{i.e.}
\begin{equation}\label{eq:Sullivan}
    ep\rightarrow e (\pi^{+})^* n\rightarrow e\gamma M^{+}n~.
\end{equation}
where $(\pi^{+})^*$ is an off-shell pion state.
Moreover, it has been shown that, when the momentum transfer between incoming and outgoing nucleon states is small, say $|t|\lesssim 0.6~\textrm{GeV}^{2}$, the properties of the slightly virtual photon do not differ appreciably from those of an actual pion \cite{Sullivan:1971kd,Qin:2017lcd}. Thus the process at hand, the \textit{Sullivan process}, provides a window to Compton scattering on pions.

Importantly enough, two contributions can be envisaged for such process: the Bethe-Heitler scattering process, where the outgoing photon is emitted by the final-state electron (right-most panel of Fig.~\ref{fig:kin}); and another contribution where the final-state photon is emitted by the outgoing meson. This latter contribution is the one of interest here. Two cases can be considered: When the scattering process takes place through the small-$t_\pi$ (see Fig.~\ref{fig:kin}) channel, giving rise to the well known deeply virtual Compton scattering on a pion whose amplitude factorises at all orders of perturbation theory \cite{Radyushkin:1997ki,Collins:1998be,Ji:1998xh} and which interferes with the Bethe-Heitler process; and when it occurs in the small-$u_\pi$ channel: DVCS in backward kinematics, bDVCS (middle panel of Fig.~\ref{fig:kin}). 

This latter case is the one studied here, with its specificities: there is no explicit proof of factorisation, but we expect the amplitude to factorise in the same way as in the case of deep exclusive meson production. Consequently, we will focus on the longitudinally polarised differential cross-section $\textrm{d}\sigma_L$. Moreover, in the small-$u_{\pi}$ region, the contribution from the Bethe-Heitler process can be neglected.

\subsection{Kinematics of backward DVCS in the Sullivan process}\label{Sec_Kinematics}

Let us start going through the kinematics of the reaction
\begin{equation}
  \label{proc}
e(l) + p(p) \to e(l') + \gamma(q') + M^+(p'_M) + n(p') \,,
\end{equation}
where $M$ is either a pseudoscalar or a vector meson, and follow the notation of \cite{Amrath:2008vx}. We denote by $m_{N}$ and $m_{M}$  the nucleon and final meson  masses, respectively; we neglect the lepton mass throughout our work. With $q = l-l' \,$ and $p_\pi = p-p'\,$, the virtual photon and virtual $\pi$ meson momenta, we define the Lorentz scalars
\begin{align}
Q^2 &= -q^2 \,, &  W^2 &= (p+q)^2 \,, &  s &= (p+l)^2 \,,     & t = (p-p')^2 \,,
\end{align}
and the energy fractions
\begin{align}
x_B &= \frac{Q^2}{2p\cdot q} \,, &
y   &= \frac{p\cdot q}{p\cdot l} \,, &
x_\pi &= \frac{p_\pi \cdot l}{p \cdot l}\,.
\end{align}
In particular, $x_\pi$ is the fraction of energy that the virtual pion takes away from the proton in the $ep$ \textit{c.m.} frame.

The momentum transfer $t$ is subject to the usual kinematic limit
\begin{equation}
  \label{t-min}
-t \ge -t_0 = \frac{x_\pi^2 m_{N}^2}{1-x_\pi} \,.
\end{equation}

We are primarily interested in the pion-target process
\begin{equation}
  \label{piproc}
e(l) + \pi^+(p_\pi) \to e(l') + \gamma(q') + M^+(p'_M)  \,.
\end{equation}
and we define the subprocess variables
\begin{align}
s_\pi   &= (p_\pi + q)^2 \,, &
u_{\pi} &= (p_{\pi}-p_{M}')^{2}\,, &
x_B^\pi &= \frac{Q^2}{2 p_\pi \cdot q} \,, &
y_\pi   &= \frac{p_\pi \cdot q}{p_\pi \cdot l} \,.
\end{align}

In the Bjorken limit, we have
\begin{align}
  \label{xBpi}
x_B^\pi    &\approx \frac{x_B}{x_\pi} \,, &
y_\pi      &= y \frac{x_B}{ x_\pi  x_B^\pi} \approx y \,, &
x_\pi y &\approx \frac{s_\pi+Q^2}{s}\,.
\end{align}

With all these kinematic variables, a description of the process can be achieved (see \textit{e.g.} \cite{Amrath:2008vx}).

\subsection{\texorpdfstring{The $e p \to e\gamma M n$ cross-section}{The ep->egamma M n}}

We would like to assess the cross-section for bDVCS on a pion, as given through the Sullivan process. Assuming the one-pion exchange approximation to hold the process' cross-section can be written as (see App.~\ref{app:CSdetails}):

\begin{equation}
    \frac{d^{2}\sigma^{ep\rightarrow e\gamma M n}}{dt dx_{\pi}}=x_{\pi}\frac{g^{2}_{\pi NN}}{8\pi^{2}}\frac{-t}{(t-m_{\pi}^{2})^{2}}F^{2}(t;\Lambda)d\sigma^{e\pi\rightarrow e\gamma M}
\end{equation}
where we followed \cite{Amrath:2008vx} and introduced a phenomenological factor $F(t;\Lambda)$, Eq.~\eqref{eq:PiNSmoothing}, smoothing the pion-nucleon vertex as the virtuality of the exchanged pion increases. Assuming factorization for the $e\pi\rightarrow e\gamma M$ subprocess and working in Bjorken kinematics, the cross-section reads, App.~\ref{app:CSdetails}
\begin{equation}\label{eq:SullivanCrossSection}
 \frac{d^{5}\sigma^{ep\rightarrow e\gamma n}}{dtdx_{\pi}dQ^{2}dx_{B} du_{\pi}}\simeq x_{\pi}\frac{g_{\pi NN}^{2}}{8\pi^{2}}\frac{-t}{(t-m_{\pi}^{2})^{2}}F^{2}(t;\Lambda)\left(\frac{\alpha_{\textrm{QED}}}{Q^{2}}\right)^{3}\frac{x_{B}^{\pi}}{x_{\pi}}(1-y)|\mathcal{A}^{M}_{L}(\xi,u_{\pi},Q)|^{2},
\end{equation}
where $\xi\simeq x_{B}^{\pi}/(2-x_{B}^{\pi})$ and the dynamics of the process is entirely encoded in the $\gamma\pi\rightarrow \gamma M$ amplitude, $\mathcal{A}_{L}^{M}$:
\begin{equation}\label{eq:Amplitudes}
    \begin{array}{rcl}
        \displaystyle \mathcal{A}_{L}^{\pi}(\xi,u_{\pi},Q) & \displaystyle = &\displaystyle \frac{16\pi\alpha_s(Q)}{9Q} \int dx dz C^{ud}_{F}(x,z,\xi) \Phi^{\pi}(z,Q)A^{\pi^+\gamma} (x,\xi,u_{\pi},Q)\\
        \\
        \displaystyle \mathcal{A}_{L}^{\rho}(\xi,u_{\pi},Q) & \displaystyle = &\displaystyle  \frac{16\pi\alpha_s(Q)}{9Q} \int dx dz C^{ud}_{F}(x,z,\xi) \Phi^{\rho}(z,Q)V^{\pi^+\gamma} (x,\xi,u_{\pi},Q)  
    \end{array}
\end{equation}
for (light) outgoing pseudoscalar and vector mesons, respectively. There $\Phi^{M}$ represents the outgoing-meson distribution amplitude and $C_{F}$ the hard-scattering kernel which, at leading order in the strong coupling, reads \cite{Collins:1996fb,Diehl:2003ny}
\begin{equation}
  \label{CF}
C^{ud}_{F}(x,z,\xi) = \frac{1}{1-z}\frac{e_{u}}{\xi-x-i\varepsilon} - \frac{1}{z}\frac{e_{d}}{\xi+x-i\varepsilon}.
\end{equation}

\begin{figure}[t]
\centering
 	\begin{subfigure}[b]{0.45\textwidth}
        \centering
        \includegraphics[scale=0.9]{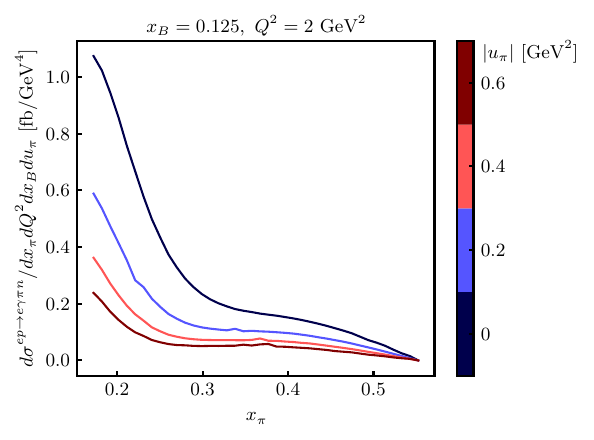}
    \caption{$ep\rightarrow e\gamma\pi n$}
    \end{subfigure} \hfill
    \begin{subfigure}[b]{0.45\textwidth}
        \centering
        \includegraphics[scale=0.9]{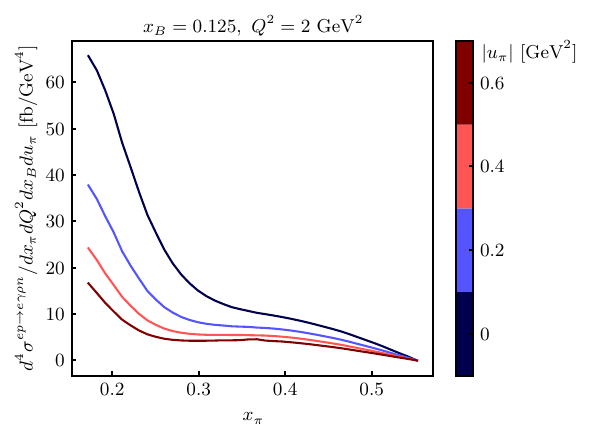}
    \caption{$ep\rightarrow e\gamma\rho n$}
    \end{subfigure}\\
\caption{\label{fig:CrossSection} Four-differential cross-section for $ep\rightarrow e\gamma\pi n$ (\textsc{Left-panel}) and $ep\rightarrow e\gamma\rho n$ (\textsc{Right-panel}) at typical JLab22 \cite{Accardi:2023chb} kinematics, $Q^{2}=2~\textrm{GeV}^{2}$, $x_{B}=0.125$ and covering the entire $x_{\pi}$ range available. Results are shown for a number of $u_{\pi}$ values.}
\end{figure}

It now becomes plain that, combining Eq.~\eqref{eq:SullivanCrossSection} and Eq.~\eqref{eq:Amplitudes}, a quantitative assessment of meson electroproduction in backward kinematics can be achieved from a proper representation of $\pi-\gamma$ TDAs. We can thus take advantage of the models developed in preceding sections to study the feasibility of such a measurement. We focus here on fixed-target electron-proton experiments as those developed in JLab. To this end we can consider typical photon virtualities of $Q^{2}=2~\textrm{GeV}^{2}$ (together with $s_{\pi}=4~\textrm{GeV}^{2}$, ensuring the proper interpretation of the process in the deeply virtual regime \cite{Amrath:2008vx}) and a number of $u_{\pi}$ values. It thus remains to fix the $x^{\pi}_{B}$ variable, \textit{i.e.} the skewness. As showed in App.~\ref{app:CSdetails}, this is strongly tight by the available beam energies. We consider $22~\textrm{GeV}$ electrons, as proposed for the anticipated upgrade \cite{Accardi:2023chb}, leaving
\begin{align}
0.17 &\lesssim x_{\pi}\lesssim 0.5\,, & 0.05 &\lesssim x_{B}\lesssim x_{\pi}\,, & 0.05 &\lesssim \xi.
\end{align}
as the available phase-space. For illustration, we choose $x_{B}=0.125$ as a representative value and, covering the available $x_{\pi}$ range, we evaluate the amplitudes Eq.~\eqref{eq:Amplitudes}. Notice that, the DA of the pion and the $\rho$-meson being symmetric in $z \to (1-z)$, the $z-$integration in Eq.~\eqref{eq:Amplitudes} factorizes in a prefactor $\int dz \Phi^\pi(z)/z$. For the sake of simplicity, we consider the asymptotic distribution amplitude, $\phi_{\rho,\pi}(z)=6f_{\pi,\rho}z(1-z)$.
Notice however that, again because $\pi$ and $\rho$ leading-twist DAs are symmetric, assessing the influence of a change in the DAs --say using different models-- is straightforward at leading order in $\alpha_{s}$, resulting in a pure multiplicative factor. We then evaluate the resulting expressions through a combination of a $25$-points modified Clenshawn-Curtis and a $15$-points Gauss-Kronrod integration rule, as implemented in the \texttt{qawc} routine of the \texttt{GNU} scientific library. As a result, we obtain estimations on the cross-section for $\pi$ and $\rho$ (deeply virtual) electroproduction in backward kinematics.

Fig.~\ref{fig:CrossSection} displays our estimation for the cross-section for deep-virtual electroproduction of $\pi$ and $\rho$ mesons, in backward kinematics, and in a typical kinematic range accessible at JLab with $22~\textrm{GeV}$ electron-beams. We show results at a number of $u_{\pi}$ values. Notably, the result for vector-meson electroproduction is appreciably larger than that for pions, by roughly factor 50. The overshooting of the form factor by a factor $3/2$ in Eq.~\eqref{eq:PDG2024FormFactors} explain only a factor 2 at the level of the cross-section. The ratio between $f_\pi$ and $f_\rho$ could explain also a factor 2.5 at the level of the cross-section. It remains roughly a factor 10 that comes from the modelling strategy and that could be tested experimentally. Compared to the forward Sullivan DVCS case, the total cross-section of backward DVCS remains two orders of magnitudes below. However, this does not preclude an experimental discovery of a signal at current and future high precision, high luminosity facilities, as the photon is expected to be measured in an area with much less background: the competitive Bethe-Heitler process is for instance negligible. Furthermore, the possibility of exploring lower-$x_{B}$ regions is not excluded \cite{Accardi:2023chb} which would greatly help in increasing the cross-sections expected.

\section{Conclusion}

We have presented a novel technique for the modeling of $\pi - \gamma$ TDAs based on a combination of the overlap representation and the covariant extension methods; strategies primarily developed for GPDs. Using existing models for the pion valence LFWFs and after developing a parametrization for those of the photon, we obtained DGLAP TDAs which produce good results for axial transition form factors and acceptable ones for the vector transition form factors. We extended these models to their ERBL region and explored the effect of scale evolution. Finally, we exploited these models in a description of backward DVCS on a pion, in the framework of the Sullivan process.

We have thus demonstrated the interest of studying the backward DVCS on a pion in the framework of the Sullivan description. The two processes discussed here are complementary. The $\gamma^{\ast} \pi \rightarrow \gamma \pi $ reaction probes the axial $\pi - \gamma$ TDA, while the $\gamma^{\ast} \pi \rightarrow \gamma\rho$ reaction probes the vector one. Both reactions are dominated by the longitudinal cross-section. Moreover, the second one selects a longitudinally polarized $\rho$ meson, a prediction one may test by studying the angular distribution of the $\rho$ meson decay. We performed our analysis at leading order in the strong coupling $\alpha_s$; NLO corrections which are known for the coefficient function \cite{Muller:2013jur, Duplancic:2016bge} will be straightforward to implement in the near future. An estimate of higher twist corrections would be most welcome but this represents now a challenge. Note that the recent attempts \cite{Braun:2014sta,Braun:2022qly,  Martinez-Fernandez:2025gub} to take into account of kinematic higher twist corrections to forward Compton processes (or related processes \cite{ Lorce:2022tiq, Pire:2023kng,Pire:2023ztb}), following the conformal operator product expansion techniques \cite{Braun:2011dg} cannot  easily be applied to backward DVCS since in this case the starting point is not the Compton tensor. Genuine higher twist corrections are neither easy to estimate.

Our study opens the way to a new access to $\pi - \gamma$ TDAs in the backward DVCS channels; our cross-section estimates allow us to hope for a first signal in JLab experiments at present energies, and for a fruitful program with dedicated experiments at higher energies, although more precise feasibility studies are clearly needed. An interesting outcome of a first experimental campaign would be the invalidation that the $\rho$ production is enhanced compared to the pion one.

As a final remark, let us point out that the very related process of backward timelike Compton scattering $\gamma p \rightarrow \gamma^{\ast} M n$ at large timelike $Q^2$ should also be addressed within the Sullivan picture to probe the crossed $\gamma - \pi$ TDA and the new $\gamma - \rho$ TDAs.

\acknowledgments
We acknowledge useful discussions with Valerio Bertone, Maxime Defurne, Wenliang Li, Kirill Semenov-Tian-Shansky and Lech Szymanowski.
This research was funded in part by l’Agence Nationale de la Recherche (ANR), project ANR-23-CE31-0019, P2IO LabEx (ANR10-LABX-0038) in the framework of Investissements d’Avenir (ANR-11-IDEX-0003-01), by the Coordenação de Aperfeiçoamento de Pessoal de Nível Superior - Brasil (CAPES) – Finance Code 001, and by Ministerio de Ciencia, Innovación y Universidades, Spain (PID2023-151418NB-I00).
For the purpose of open access, the authors have applied a CC-BY public copyright licence to any Author Accepted Manuscript (AAM) version arising from this submission.

\pagebreak
\appendix

\section{Details on cross-section computation}\label{app:CSdetails}

In this appendix, we sketch some details on the derivation of the cross-section for $ep\rightarrow e\gamma M n$ process. The interesting reader can find more details in \cite{Amrath:2008vx,Amrath:2007zz}. The starting point is the standard cross-section formula,
\begin{equation}\label{eq:SullivanCS}
d\sigma^{ep\rightarrow e\gamma Mn}=\frac{1}{F_{ep}}\sum_{\textrm{Pol}.}|\mathcal{M}|^{2}d\Pi_{4},
\end{equation}
where $F_{ep}=\sqrt{\lambda(s,m_{N}^{2},0)}=2(s-m_{N}^{2})$ is the electron-proton flux-factor, $\mathcal{M}$ is the process' amplitude and $d\Pi_{4}$ is the Lorentz-invariant four-body phase-space element. Notice that $\sum_{\textrm{Pol.}}$ denotes here sum over the final state polarization and averaging over the initial-state ones.

In the one-pion-exchange approximation, the amplitude $\mathcal{M}$ can be decomposed as
\begin{equation}\label{eq:AmplitudeSullivanOPE}
\mathcal{M}=\bar{u}_{\sigma'}(p')i\gamma_{5}\sqrt{2}g_{\pi NN}u_{\sigma}(p)\frac{i}{t-m_{\pi}^{2}}F(t;\Lambda)\mathcal{M}_{e\pi^{+}\rightarrow e\gamma M^{+}},
\end{equation}
where $u_{\sigma}(p)$ is the nucleon spinor with momentum $p$ and polarization $\sigma$; $i\gamma_{5}\sqrt{2}g_{\pi NN}$ is the pion-nucleon vertex whose strength is measured by $g_{\pi NN}=13.05$ \cite{Stoks:1992ja}, $i/(t-m_{\pi}^{2})$ is the pion propagator and, $F(t;\Lambda)$ is a phenomenological factor\footnote{Different choices for such a ``form-factor'' can be made. A discussion on the subject can be found in \cite{McKenney:2015xis}.} introduced to smooth the pion-nucleon vertex as the offshelness of the exchanged photon increases \cite{Amrath:2008vx}
\begin{equation}\label{eq:PiNSmoothing}
F(t;\Lambda=800~\textrm{MeV})=\frac{\Lambda^{2}-m_{\pi}^{2}}{\Lambda^{2}-t}.
\end{equation}

Combining Eq.~\eqref{eq:SullivanCS} and Eq.~\eqref{eq:AmplitudeSullivanOPE} we readily obtain
\begin{equation}
\begin{array}{rcl}
\displaystyle d\sigma^{ep\rightarrow e\gamma M n} &\displaystyle = &\displaystyle -\frac{g_{\pi NN}^{2}}{F_{ep}}\frac{F^{2}(t;\Lambda)}{(t-m_{\pi}^{2})^{2}}\sum_{\sigma,\sigma'}u_{\sigma'}(p')\bar{u}_{\sigma'}(p')\gamma_{5}u_{\sigma}(p)\bar{u}_{\sigma}(p)\gamma_{5}(p)\sum_{\textrm{Pol.}}|\mathcal{M}_{e\pi^{+}\rightarrow e\gamma M^{+}}|^{2}d\Pi_{4}\\
\\
&\displaystyle = &\displaystyle 2\frac{g_{\pi NN}^{2}}{F_{ep}}\frac{-t}{(t-m_{\pi}^{2})^{2}}F^{2}(t;\Lambda)\sum_{\textrm{Pol.}}|\mathcal{M}_{e\pi^{+}\rightarrow e\gamma M^{+}}|^{2}d\Pi_{4},
\end{array}
\end{equation}
which, defining the electron-pion flux-factor as
\begin{equation}
    F_{e\pi}=2(W^{2}-m_{\pi}^{2})=2(t-m_{\pi}^{2})+4(p_{\pi}\cdot l)\simeq 4(p_{\pi}\cdot l)=4x_{\pi}(p\cdot l)=x_{\pi}2(s-m_{N}^{2})=x_{\pi}F_{ep},
\end{equation}
and writing the four-body phase-space element in terms of the three-body one \cite{MorgadoChavez:2022men} can be cast as
\begin{equation}
    d\sigma^{ep\rightarrow e\gamma M n}=x_{\pi}\frac{g_{\pi NN}^{2}}{16\pi^{3}}\frac{-t}{(t-m_{\pi}^{2})^{2}}F^{2}(t;\Lambda)dtdx_{\pi}d\psi_{n}d\sigma^{e\pi^{+}\rightarrow e\gamma M^{+}},
\end{equation}
with
\begin{equation}
d\sigma^{e\pi^{+}\rightarrow e\gamma M^{+}}=\frac{1}{F_{e\pi}}\sum_{\textrm{Pol.}}|\mathcal{M}_{e\pi^{+}\rightarrow e\gamma M^{+}}|^{2}d\Pi_{3},
\end{equation}
the cross-section for the $e\pi^{+}\rightarrow e\gamma M^{+}$ sub-process, defined in analogy to Eq.~\eqref{eq:SullivanCS}. Notice that the above way of writing explicitly decouples the ``auxiliary'' process where a pion is emitted by a nucleon from the actually interesting one, where the probing electron beam scatters off such pion.

If as discussed in Sec.~\ref{sec:Amplitude} we focus on the contribution from longitudinally polarized photons to the cross-section for $e\pi^{+}\rightarrow e\gamma M^{+}$ we find:
\begin{equation}
    d\sigma^{e\pi^{+}\rightarrow e\gamma M^{+}}=\frac{1}{F_{e\pi}}\left(\frac{2e^{3}}{Q}\right)^{2}\frac{1-y_{\pi}}{y_{\pi}^{2}}|\mathcal{A}^{\gamma^{\ast}\pi}_{L}(\xi,u_{\pi},Q)|^{2}d\Pi_{3}(l,p_{\pi}=p-p').
\end{equation}

The last element to be included is thus the three-body phase-space \cite{MorgadoChavez:2022men}, allowing us to write
\begin{equation}
\displaystyle \frac{d^{5}\sigma^{e\pi^{+}\rightarrow e\gamma M^{+}}}{dQ^{2}dx_{B}d\psi_{e}du_{\pi}d\psi} \simeq \frac{1}{(2\pi)^{2}}\left(\frac{\alpha_{\textrm{QED}}}{Q^{2}}\right)^{3}\frac{y^{2}x_{B}^{\pi}}{x_{\pi}\sqrt{1+\epsilon^{2}}}\frac{1-y}{y^{2}}|\mathcal{A}^{M}_{L}(\xi,u_{\pi},Q)|^{2},
\end{equation}
where we assumed Bjorken kinematics, Eq.~\eqref{xBpi}. Finally, if we integrate over the azimuthal angles $\psi$ and $\psi_{e}$ getting an extra factor $(2\pi)^{2}$ which yields
\begin{equation}
\displaystyle \frac{d^{3}\sigma^{e\pi^{+}\rightarrow e\gamma M^{+}}}{dQ^{2}dx_{B}du_{\pi}} = \left(\frac{\alpha_{\textrm{QED}}}{Q^{2}}\right)^{3}\frac{y^{2}x_{B}^{\pi}}{x_{\pi}\sqrt{1+\epsilon^{2}}}\frac{1-y}{y^{2}}|\mathcal{A}^{M}_{L}(\xi,u_{\pi},Q)|^{2},
\end{equation}
which, notice, coincides with the result reported in \cite{Diehl:2003ny} (Eq.~(376)) for the contribution given by longitudinally-polarized photons to the cross-section for $ep\rightarrow epM$.

Finally, arranging everything together we get the cross-section for the Sullivan process
\begin{equation}\label{eq:SullivanCSres}
    \frac{d^{6}\sigma^{ep\rightarrow e\gamma M n}}{dtdx_{\pi}d\psi_{n}dQ^{2}dx_{B} du_{\pi}}=x_{\pi}\frac{g_{\pi NN}^{2}}{16\pi^{3}}\frac{-t}{(t-m_{\pi}^{2})^{2}}F^{2}(t;\Lambda)\left(\frac{\alpha_{\textrm{QED}}}{Q^{2}}\right)^{3}\frac{y^{2}x_{B}^{\pi}}{x_{\pi}\sqrt{1+\epsilon^{2}}}\frac{1-y}{y^{2}}|\mathcal{A}^{M}_{L}(\xi,u_{\pi},Q)|^{2},
\end{equation}
which, after integration on the azimuthal angle of the scattered neutron, $\psi_{n}$ gives:
\begin{equation}\label{eq:Cross-section}
    \frac{d^{5}\sigma^{ep\rightarrow e\gamma M n}}{dtdx_{\pi}dQ^{2}dx_{B} du_{\pi}}=x_{\pi}\frac{g_{\pi NN}^{2}}{8\pi^{2}}\frac{-t}{(t-m_{\pi}^{2})^{2}}F^{2}(t;\Lambda)\left(\frac{\alpha_{\textrm{QED}}}{Q^{2}}\right)^{3}\frac{y^{2}x_{B}^{\pi}}{x_{\pi}\sqrt{1+\epsilon^{2}}}\frac{1-y}{y^{2}}|\mathcal{A}^{M}_{L}(\xi,u_{\pi},Q)|^{2},
\end{equation}

\section{JLab kinematics}

In this work we are working towards a first assessment of the cross-section for $\pi$- and $\rho$- mesons electroproduction off a neutron in backward kinematics at JLab. Having developed a consistent model for $\pi^{+}\gamma$ it is time to concentrate on the kinematics available to process. In this context, a number of kinematic constraints arise on an experimental as well as a theoretical basis; in the following we sketch them, setting the ground for a numerical evaluation of the cross-section Eq.~\eqref{eq:Cross-section}

\begin{itemize}
    \item $s_{ep}$: The Continuous Electron Beam Accelerator Facility (CEBAF) at JLab develops a scientific program based on fixed-target experiments. Up to date available beam energies reach $12~\textrm{GeV}$ but an upgrade to $22~\textrm{GeV}$ electron-beam energies is planned \cite{Accardi:2023chb}. In practice, this fixes the electron-proton center of mass energy to:
    \begin{equation}
        s^{(12)}_{ep}\simeq 23~\textrm{GeV}^{2},\qquad\textrm{JLab12},
    \end{equation}
    or after the upgrade,
    \begin{equation}
        s^{(22)}_{ep}\simeq 42~\textrm{GeV}^{2},\qquad\textrm{JLab22}.
    \end{equation}
    
    \item $s_{\pi}$: For the final-state meson and photon to be properly detected, a minimum energy is required. We impose here a constraint \cite{Amrath:2008vx}
    \begin{equation}
        s_{\pi} \geq s_{\pi}^{\textrm{Min.}}=4~\textrm{GeV}^{2}.
    \end{equation}
    
    \item Mandelstam $t$: Ensuring that the proton-neutron transition is dominated by the emission of a single pion requires cutting on the squared momentum transfer between those such states. Previous works suggest a constraint on the maximum allowed value of $|t|\leq 0.6~\textrm{GeV}^{2}$ \cite{Qin:2017lcd,Chavez:2021koz,Amrath:2008vx}. Further, a lower limit given by the nucleon mass exist for such variable:
    \begin{equation}
        \frac{x_{\pi}^{2}m_{N}^{2}}{1-x_{\pi}}\leq |t|\leq 0.6~\textrm{GeV}^{2}
    \end{equation}
    
    \item $Q^{2}$: To guarantee the description of the scattering process in Bjorken kinematics and justify the theoretical development of the paper, a lower bound on the allowed $Q^{2}$ values is imposed:
    \begin{equation}
        Q^{2}\geq Q^{2}_{\textrm{Min.}}=2~\textrm{GeV}^{2}.
    \end{equation}
    
    \item $y$: The variable $y$ can be interpreted as a measure of the energy transfer from the electron beam to probing photon. Usually, in order to suppress QED radiative corrections, an upper bound on $y$ is imposed \cite{Amrath:2008vx}. We choose a mild-value,
    \begin{equation}
        y\leq y_{\textrm{Max.}}=0.85.
    \end{equation}
    \item $x_{\pi}$: The interpretation of the process in the one-pion-exchange approximation also impose constraints on the kinematic variable $x_{\pi}$, which measures the energy transferred from the target proton to the virtual pion:
    \begin{equation}
          x_{\pi}y\simeq x_{\pi}\leq \frac{-|t_{\textrm{Max.}}|\pm\sqrt{|t|^{2}+4m_{N}^{2}|t_{\textrm{Max.}}|}}{2m_{N}^{2}}\simeq 0.55
    \end{equation}
    
    In addition, using $x_{\pi}y\simeq x_{\pi}y_{\pi}$, a lower bound for $x_{\pi}$ can be found. Indeed,
    \begin{equation}
        x_{\pi}y_{\pi}=\frac{2p_{\pi}\cdot q}{2p\cdot l}=\frac{s_{\pi}-t+Q^{2}}{s_{ep}-m_{N}^{2}}\simeq\frac{s_{\pi}+Q^{2}}{s_{ep}-m_{N}^{2}},
    \end{equation}
    
    and therefore
    \begin{equation}
        \frac{1}{y_{\textrm{Max.}}}\frac{s_{\pi}+Q^{2}}{s_{ep}-m_{N}^{2}}\lesssim x_{\pi}\leq \frac{-|t_{\textrm{Max.}}|\pm\sqrt{|t|^{2}+4m_{N}^{2}|t_{\textrm{Max.}}|}}{2m_{N}^{2}}\simeq 0.55
    \end{equation}
    
    \item $x_{B}$: The standard Bjorken variable is also constrained by the available kinematic configuration. Indeed, 
    \begin{equation}
        x_{B}=\frac{Q^{2}}{2p\cdot q}=\frac{y_{\pi}x_{\pi}}{y}\frac{Q^{2}}{2 p_{\pi}\cdot q}= \frac{y_{\pi}x_{\pi}}{y}\frac{Q^{2}}{s_{\pi}+Q^{2}-t}\simeq x_{\pi}\frac{Q^{2}}{s_{\pi}+Q^{2}},
    \end{equation}
    which, together with the lower bound on $x_{\pi}$ yields
    \begin{equation}
     \frac{1}{y_{\textrm{Max.}}}\frac{Q^{2}}{s_{ep}-m_{N}^{2}}\leq x_{B}\lesssim x_{\pi};
    \end{equation}
    where the upper bound arises from the constraint $\xi<1$.
\end{itemize}

In line with these kinematic constraints we decide to explore the behavior of the cross-section at fixed $Q^{2}=2~\textrm{GeV}^{2}$ and selected values of $u_{\pi}$. We take a glimpse at the effect of varying $x_{B}$ choosing representative values $x_{B}=0.2,0.3$, accessible at both Jab$12$ and JLab$22$. Notice also that, within this kinematic configuration, the $t$-dependence of the $e\pi^{+}\rightarrow e\gamma M^{+}$ cross-section can be neglected, \textit{i.e.} if we drop terms of order $t/Q^{2}$,
\begin{equation}
    \frac{1}{\sqrt{1+\epsilon^{2}}}=\frac{1}{\sqrt{1+4t\frac{(x_{B}^{\pi})^{2}}{Q^{2}}}}=1-2(x_{B}^{\pi})^{2}\frac{t}{Q^{2}}+\mathcal{O}\left(\frac{t^{2}}{Q^{4}}\right)
\end{equation}
which, in the less favorable scenario, introduces a $40\%$. We may then drop such term and write the cross-section as
\begin{equation}
 \frac{d^{5}\sigma^{ep\rightarrow e\gamma Mn}}{dtdx_{\pi}dQ^{2}dx_{B} du_{\pi}}\simeq x_{\pi}\frac{g_{\pi NN}^{2}}{8\pi^{2}}\frac{-t}{(t-m_{\pi}^{2})^{2}}F^{2}(t;\Lambda)\left(\frac{\alpha_{\textrm{QED}}}{Q^{2}}\right)^{3}\frac{y^{2}x_{B}^{\pi}}{x_{\pi}}\frac{1-y}{y^{2}}|\mathcal{A}^{M}_{L}(\xi,u_{\pi},Q)|^{2}
\end{equation}
which definitely allows for integration over $t$ to give
\begin{equation}\label{eq:CS}
    \frac{d^{4}\sigma^{ep\rightarrow e\gamma Mn}}{dx_{\pi}dQ^{2}dx_{B}du_{\pi}}=\Pi\left(x_{\pi},|t|_{\textrm{Max.}}\right)\left(\frac{\alpha_{\textrm{QED}}}{Q^{2}}\right)^{3}\frac{y^{2}x_{B}^{\pi}}{x_{\pi}}\frac{1-y}{y^{2}}|\mathcal{A}^{M}_{L}(\xi,u_{\pi},Q)|^{2}
\end{equation}
with
\begin{equation}
\begin{array}{rcl}
    \displaystyle \Pi(x_{\pi},|t|_{\textrm{Max.}}) &\displaystyle = &\displaystyle x_{\pi}\frac{g_{\pi NN}}{8\pi^{2}}\int^{-|t_{0}|}_{-|t|_{\textrm{Max.}}}\frac{-t}{(t-m_{\pi})^{2}}F^{2}(t;\Lambda)dt\\
    \\
    &\displaystyle = &\displaystyle x_{\pi}\frac{g_{\pi NN}^{2}}{8\pi^{2}}\left[\frac{\Lambda^{2}+m_{\pi}^{2}}{\Lambda^{2}-m_{\pi}^{2}}\log\left(\frac{\Lambda^{2}-t}{m_{\pi}^{2}-t}\right)-\frac{\Lambda^{2}}{\Lambda^{2}-t}-\frac{m_{\pi}^{2}}{m_{\pi}^{2}-t}\right]_{-|t|_{\textrm{Max}}}^{-|t_{0}|},
\end{array}
\end{equation}
as a measure of the pion's flux into the scattering process $e\pi\rightarrow e\gamma M$ \cite{Amrath:2008vx}.

With this construction we write the cross-section formula as
\begin{equation}
    \frac{d^{4}\sigma^{ep\rightarrow e\gamma Mn}}{dx_{\pi}dQ^{2}dx_{B}du_{\pi}}=\Pi(x_{\pi},|t|_{\textrm{Max.}})\left(\frac{\alpha_{\textrm{QED}}}{Q^{2}}\right)^{3}\frac{(s-m_{N}^{2})x_{B}-Q^{2}}{(s-m_{N}^{2})x_{\pi}^{2}}|\mathcal{A}^{M}(\xi,u_{\pi},Q)|^{2},
\end{equation}
where we wrote $yx_{B}=Q^{2}/(s-m_{N}^{2})$ and we note
\begin{equation}
    \xi=\frac{x_{B}}{2x_{\pi}-x_{B}}.
\end{equation}

\bibliography{SulbDVCS}

\end{document}